\documentclass[aps,pre,amsmath,amssymb,reprint]{revtex4-1}
\usepackage[utf8]{inputenc}
\usepackage{mfirstuc}
\usepackage{graphicx}
\usepackage{subfigure}
\usepackage{float}
\usepackage{gensymb}
\usepackage{amsmath}
\usepackage{color}
\usepackage{multirow}
\usepackage{comment}
\usepackage{enumitem}
\usepackage[colorlinks=true, linkcolor=blue, citecolor=blue, urlcolor=blue]{hyperref}

\newcommand{\SK}[1]{\textcolor{black}{{#1}}}

\begin{document}

\title{Perspective: The Physics of Active Solids - From Hamiltonians to Active Matter Models}
%\title{\textcolor{red}{Perspective: The Physics of Active Solids - Coupling of Local Active Driving with Long-Wavelength Modes}}
%\title{Perspective: The Physics of Active Solids - Coupling of Local Active Driving with Long-Wavelength Phonon Modes}

\author{Antik Bhattacharya$^{1}$}
\email{abhattacharya@tifrh.res.in}
\author{J\"{u}rgen Horbach$^{2}$}
\email{horbach@thphy.uni-duesseldorf.de}
\author{Smarajit Karmakar$^{1}$}
\email{smarajit@tifrh.res.in}
\affiliation{
$^{1}$Tata Institute of Fundamental Research Hyderabad, 36/P, Gopanpally Village, Serilingampally Mandal, Ranga Reddy District, Hyderabad, Telangana 500046, India}

\affiliation{
$^{2}$Institut f\"ur Theoretische Physik II: Weiche Materie, Heinrich-Heine-Universit\"at D\"usseldorf, Universit\"atsstraße 1, 40225 D\"usseldorf, Germany}
\date{\today}

\begin{abstract}
The physics of active matter, wherein constituent particles consume energy to generate autonomous motion, has revolutionized non-equilibrium statistical mechanics. While a large body of work has successfully elucidated the behavior of dilute active systems, the dense regime - characterized by ``active glasses and active solids'' - presents profound challenges that defy conventional theoretical frameworks. Recent observations reveal two striking features in these dense systems: an apparent enhancement of Mermin-Wagner-Hohenberg (MWH) fluctuations leading to anomalous long-wavelength density fluctuations, and a remarkable correspondence between activity-induced annealing and annealing via oscillatory shear. In this perspective article, we propose a novel approach toward a deeper understanding of dense active matter: by developing active Hamiltonian models as equilibrium reference frameworks, we map out pathways toward non-equilibrium active systems. This strategy allows us to elucidate both the correspondence between driven and active systems and the enhanced MWH fluctuations, which likely arise from a strong coupling between spatially random active forces and long-wavelength density (phonon) modes. We outline a comprehensive roadmap employing complementary approaches, including the active Hamiltonian formalism, comparative studies of oscillatory shear in active and passive solids, and investigations of chiral active matter. Establishing this activity-oscillatory shear correspondence across diverse systems is essential to demonstrate its universality, reveal the underlying large-scale emergent physics, and place our hypothesis on a firmer theoretical ground.
\end{abstract}

\maketitle

\noindent{\large\bf Introduction:}
\SK{The field of active matter has emerged as one of the most dynamic and rapidly evolving areas of condensed matter and soft matter physics \cite{ramaswamy2010mechanics, marchetti2013hydrodynamics, palacci2013living, vicsek1995novel}. Unlike passive equilibrium systems, where dynamics are driven solely by thermal fluctuations, active matter consists of self-driven constituents. These building blocks—be they birds in a flock, bacteria in a colony, or synthetic Janus colloids—continuously consume energy at the microscopic scale to generate self-propulsion. This constant injection of energy at the individual particle level drives the system fundamentally out of equilibrium, leading to rich collective behaviors that have no counterpart in passive systems \cite{ballerini2008interaction, toner1998flocks}.}

\SK{The relevance of active matter extends across a vast range of length scales. In nature,} {these phenomena manifest} \SK{in the coordinated motion of fish schools, the swarming of ant colonies, and the complex dynamics of cellular cytoplasm \cite{parry2014bacterial, gravish2015glass}. Biological processes such as wound healing and cancer progression also exhibit signatures of active matter dynamics,} {with cells undergoing} \SK{collective migration and jamming transitions reminiscent of glassy systems \cite{angelini2011glass}. In the synthetic realm, active systems are realized through thermophoretic Janus colloids, vibrated granular rods, and light-activated active gels \cite{dauchot2005dynamical, dreyfus2005microscopic}.} {These experimental systems enable} \SK{the controlled study of non-equilibrium statistical mechanics, providing a testing ground for theories that aim to describe life-like motility using fundamental physical laws.}

{While the hydrodynamics of dilute active fluids (e.g., bacterial suspensions) are now well-established, active matter in the dense regime remains poorly understood. In closely packed environments, self-propulsion directly competes with steric crowding and elastic restoring forces. Crucially, dense disordered active matter exhibits central features of glassy systems, such as dynamical heterogeneity, non-exponential relaxation, and particle caging \cite{takatori2020motility}. This profound analogy has led to the emergence of the concept of ``active glasses'' \cite{mandal2016active, zhou2009universal, angelini2011glass, parry2014bacterial, berthier2013non, berthier2014nonequilibrium}.}

\begin{figure*}[!htpb]
\centering
\includegraphics[width=0.98\textwidth]{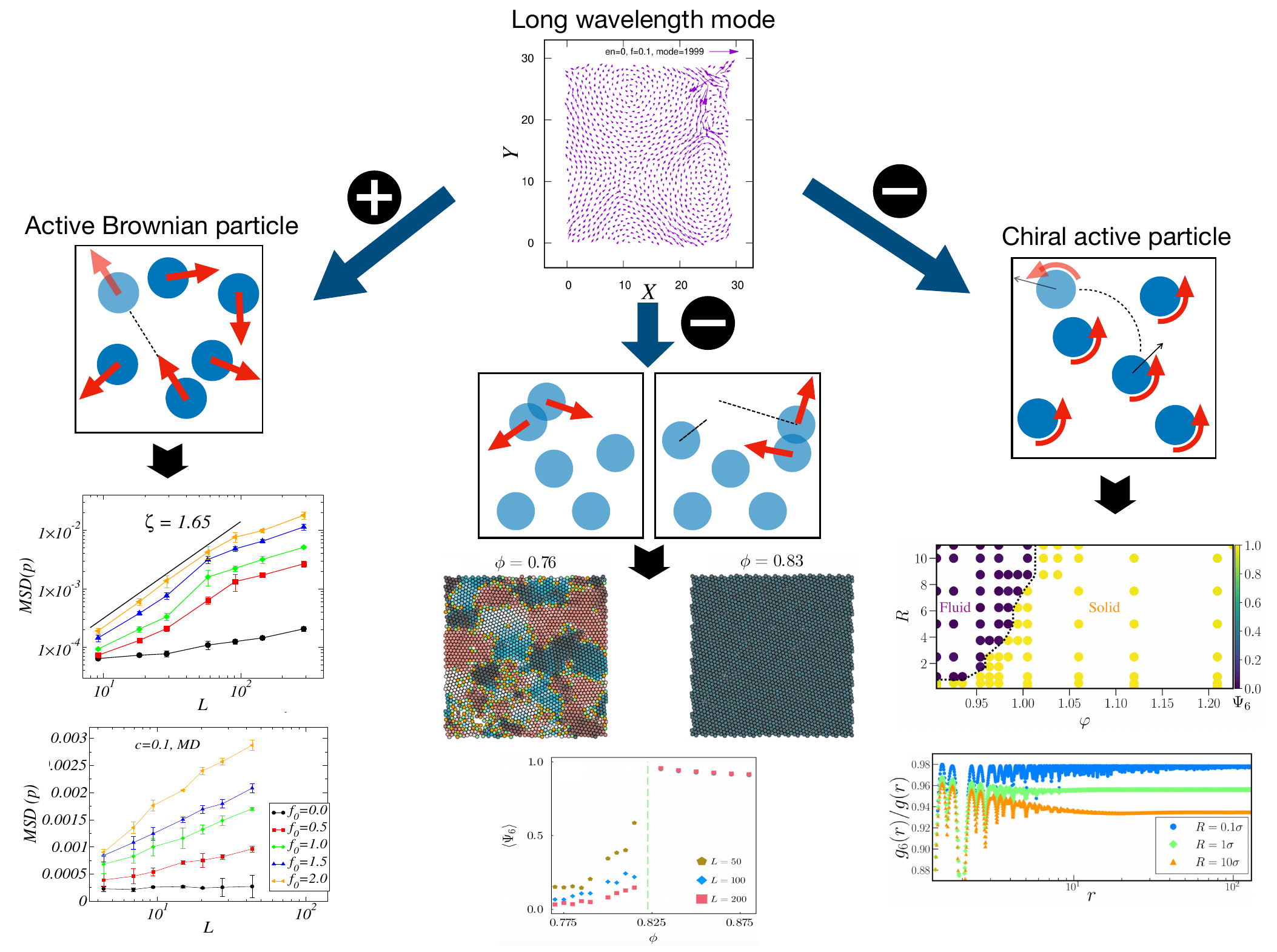}
\caption{{\bf Anomalous Mermin-Wagner-Hohenberg (MWH) fluctuations in active systems:} This schematic illustrates how positive and negative coupling between active driving and long wavelength modes affects system behavior. On the left, Active Brownian Particles (ABPs) couple positively with long wavelength modes, resulting in increased density fluctuations and destabilization of active solids even in three dimensions (3D). In the middle, we show negative coupling for another type of activity, in which particles receive a random kick whenever they come into contact, at zero temperature. This leads to strong suppression of density fluctuations and stabilization of crystalline positional order even in two dimensions (2D). On the right, chiral active particles also exhibit negative coupling, again suppressing density fluctuations and enabling stable crystalline solids. These strong couplings produce phenomena not observed in equilibrium systems.} \label{fig:1}
\end{figure*}
{The study of active glasses aims to understand glass formation far from equilibrium, utilizing activity as a control parameter to help crack the puzzle of the glass transition in passive systems. Yet, a central theoretical barrier persists: the lack of a proven statistical mechanics framework for active matter. Unlike equilibrium systems governed by the Boltzmann distribution, active systems lack a direct mapping between microscopic states and macroscopic variables. Approximating steady-state dynamics via an effective temperature ($T_{\text{eff}}$) offers only limited utility \cite{fodor2016far, loi2008effective, nandi2018random, ni2013pushing, berthier2019glassy}. Although $T_{\text{eff}}$ successfully models the velocity distribution of active Ornstein-Uhlenbeck particles in harmonic traps and two-point relaxation times in supercooled fluids, it fails for higher-order correlations. Notably, the four-point susceptibility $\chi_4(t)$—-which measures the length scale of dynamical heterogeneity—-cannot be described by a single effective temperature, signaling that collective fluctuations in active glasses decouple from standard thermal mechanisms \cite{paul2023dynamical}.}

\begin{figure*}[htpb]
  \centering
  \includegraphics[width=0.98\textwidth]{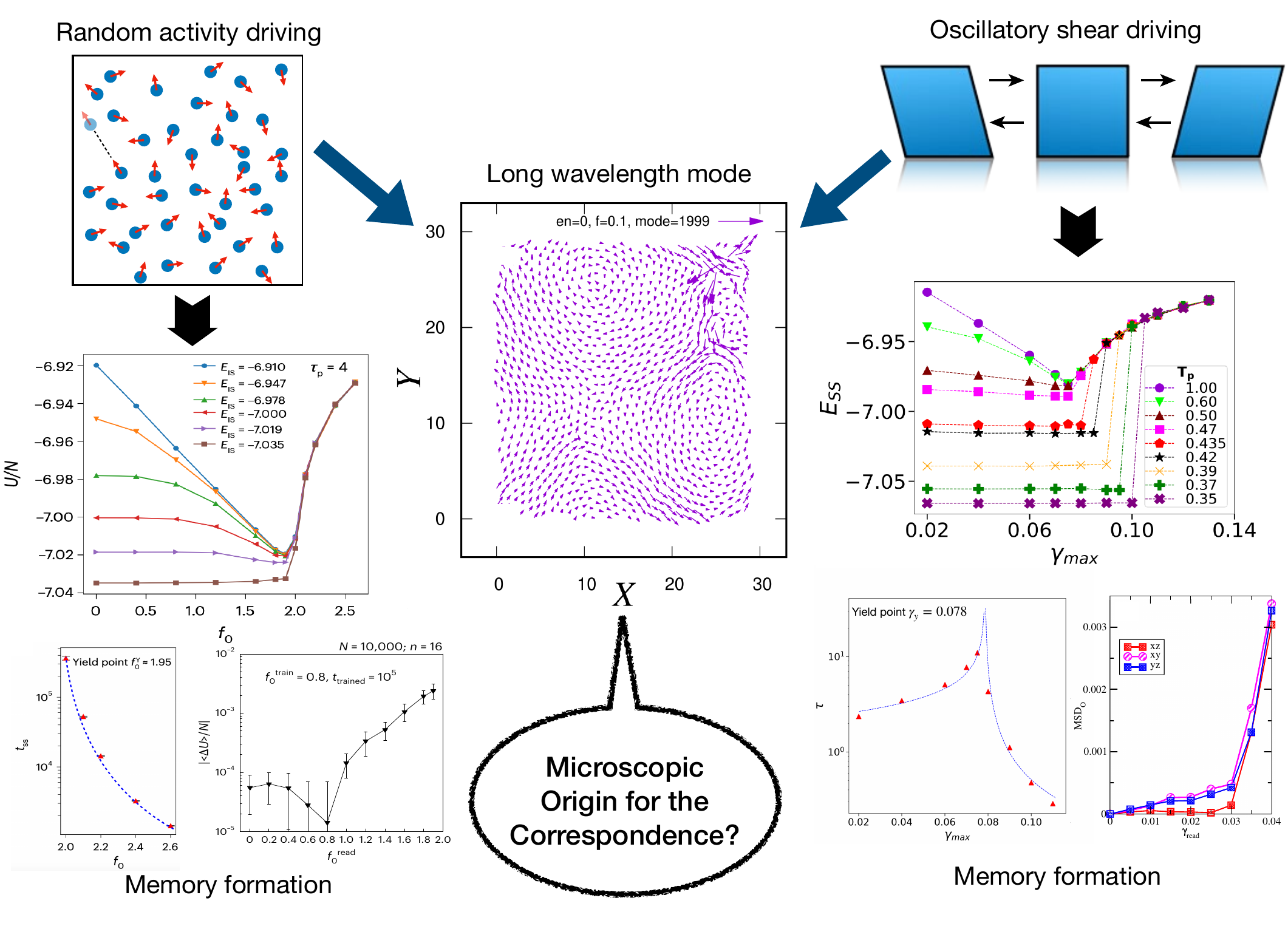}
  \caption{{\bf Activity-Oscillatory Shear Correspondence:} This schematic shows the observed equivalence between oscillatory shear response of an amorphous solids and response under active particle doping. In the left, we demonstrated the dynamical response and annealing effects obtained using active particle doping. The Energy-activity state diagram shows that annealing effects can be obtained for poorly annealed glasses, but below a critical energy, further annealing can not be achieved via active particle doping. The phenomena also shows a growth of a critical timescale and memory formation (see the text for details). On the right, we show similar dynamical responses of an amorphous solid under oscillatory shear deformation~\cite{RoniNatComm2026}. One once again observes that poorly annealed glasses can be annealed well using oscillatory shear protocol but below a critical energy, oscillatory shear can not anneal the sample further. This phenomena is also associated with a diverging timescale and memory of oscillatory shear amplitude can be encoded in this materials very efficiently. This schematic highlights that Activity-Oscillatory shear correspondence is not superficial rather it is much deeper and probably also underlies the strong coupling between active driving and long wavelength modes as discussed in the previous section.}
\label{fig:2}
\end{figure*}
\SK{This perspective is motivated by two specific, puzzling observations in active solids that suggest a deep, unexamined link between activity, elasticity, and fluctuations.}
%\begin{enumerate}
%\item{
\vskip +0.05in
%\noindent{\bf I. Violation of the Mermin-Wagner-Hohenberg (MWH) Theorem:}
\noindent{\bf I. {Enhanced Mermin-Wagner-Hohenberg (MWH) fluctuations:}}
{The Mermin–Wagner–Hohenberg (MWH) theorem states that in 2D equilibrium systems with continuous symmetry and short-range interactions, long-wavelength thermal fluctuations prevent long-range positional order at finite temperatures, causing the Debye–Waller factor to diverge logarithmically with system size $L$ \cite{hohenberg1967existence, mermin1966absence, mermin1968crystalline, jancovici1967infinite}. Active solids, however, strikingly show a different behavior. Systems of run-and-tumble particles (RTPs) exhibit ``hyper-fluctuations'' where positional variance diverges as a power law with system size \cite{dey2025enhanced}. Driven by the active force $f_{0}$, these stochastic forces couple to elastic modes, shifting the phonon dispersion from a linear scaling to $\omega(q) \sim q^\alpha$ ($\alpha \approx 1.5$). As this anomaly persists in 3D, 3D active solids become inherently unstable. This is a phenomenon unique to active matter (Fig.~\ref{fig:1}, right panel).}

{Conversely, alternative active forcing mechanisms can induce negative coupling to long-wavelength modes, suppressing fluctuations and restoring ideal positional order \cite{galliano2023two} (Fig.~\ref{fig:1}, middle panel). Chiral active particles offer a versatile platform where this coupling can be continuously tuned: positive coupling destabilizes 2D order, whereas negative coupling reinforces stabilization as chirality intensifies \cite{kuroda2025long} (Fig.~\ref{fig:1}, left panel). Hence, active driving inherently acts as a bipolar control parameter for elastic fluctuations, unlocking novel structural phases unseen in equilibrium.}

\vskip +0.05in
%\item {
\noindent{\bf II. The Activity-Oscillatory Shear Correspondence:} 
{The second focus area addresses the mechanical annealing of amorphous solids. Subjecting a glass to cyclic deformation (oscillatory shear) drives the system to lower energy states, optimizing stability~\cite{RoniNatComm2026}. Probing this effect, recent simulation studies reveal that active driving triggers an identical annealing response \cite{sharma2025activity, goswami2025yielding}. Consequently, yielding behavior and mechanical failure modes in activity-annealed glasses map directly onto shear-annealed systems: poorly annealed configurations fail via ductile necking, whereas well-annealed samples undergo brittle shear banding. Notably, both pathways exhibit a lower critical energy threshold below which annealing ceases, pointing to a unified mechanism. This analogy also governs memory encoding: amorphous solids remember the applied shear amplitude in the same manner they encode the force magnitude of embedded active dopants (Fig.~\ref{fig:2}). Since macroscopic oscillatory shear inherently couples to long-wavelength modes, the equivalent material response elicited by localized active doping strongly suggests an underlying coupling between active particle dynamics and long-wavelength elastic modes.}

{Enhanced MWH fluctuations in active systems point to a coupling of active forces to long-wavelength density modes without a passive counterpart. However, the similarity between active matter and passive systems under oscillatory shear suggests that active models might be related to Hamiltonian systems in external fields. In both cases, equilibrium frameworks break down: the stationary probability density $P(\Gamma)$ of a microstate $\Gamma$ no longer follows the Boltzmann form, $P(\Gamma) \propto \exp\left(-\beta \mathcal{H}_{\textrm{eff}}(\Gamma) \right)$, where $\beta = 1/(k_{\textrm{B}}T)$ and $\mathcal{H}_{\textrm{eff}}(\Gamma)$ denote the inverse thermal energy and (effective) Hamiltonian, respectively. Moreover, these non-equilibrium systems lack time-reversibility. For description via stochastic differential equations, this implies a violation of detailed balance and a breakdown of fluctuation-dissipation theorems.}

{A key challenge in the study of active systems is establishing a connection to Hamiltonian reference systems, from which the pathway towards non-equilibrium active states can be traced. Such a connection would make it possible to identify the underlying processes that lead, for example, to the coupling of active forces to long-wavelength density modes and, consequently, to enhanced MWH fluctuations. In this perspective article, we propose a comprehensive, multi-pronged research roadmap employing a variety of complementary theoretical, computational, and potentially experimental approaches to achieve this goal. Specifically, we introduce a Hamiltonian system that captures essential features of activity, such as the presence of self-propelling forces. From this starting point, we systematically chart the trajectories leading from the active Hamiltonian framework to generic models of active matter, such as Active Brownian Particles (ABP) and Run-and-Tumble Particles (RTP).} 

{Moreover, establishing the existence and universality of the activity-oscillatory shear correspondence across a diverse array of systems is essential. Doing so will not only validate the proposed microscopic mechanism but also reveal the universal, large-scale emergent physics governed by symmetry principles and conservation laws—transcending microscopic details. Such a demonstration would place the hypothesis on a significantly firmer theoretical ground, potentially opening pathways toward a unified theoretical description of dense active matter. The remainder of this perspective outlines how to rigorously test this hypothesis through Hamiltonian modeling, chirality studies, and possibly extending the same in systems like active gels, and patchy colloids, etc.}

\vskip +0.1in
\noindent{\large\bf Active Hamiltonian Formulation: An Effective Equilibrium Approach:}
{To systematically investigate the coupling between activity and soft modes, we require a theoretical framework that allows for precise mathematical treatment. Standard active matter models, such as ABP or RTP, are inherently dissipative and lack a defined Hamiltonian. This makes it difficult to apply powerful tools from equilibrium statistical mechanics, such as Hessian analysis (to study phonons) or free-energy landscapes. To bridge this gap, it is desirable to find a model system that possesses a well-defined Hamiltonian structure while producing collective behavior reminiscent of active matter \cite{loland2016coupling}. Fortunately, recent efforts, such as those by Casiulis {\it et al.}~\cite{casiulis2020velocity}, have made significant progress in this direction; we refer to such frameworks here as ``Active Hamiltonian'' (AH) systems.}

{The primary difficulty in analyzing active matter is the absence of a conserved energy and the non-conservative nature of self-propulsion forces. However, recent theoretical advances suggest that the phenomenological features of active matter—such as swarming—can be reproduced in equilibrium systems that feature non-trivial couplings between spatial and internal degrees of freedom. By designing a Hamiltonian system that mimics active behavior, one can utilize the comprehensive framework of statistical mechanics, including the calculation of exact phonon spectra, to uncover the origin of anomalous MWH fluctuations as well as the observed equivalence between active driving and oscillatory shear. We first focus on the model introduced by Casiulis {\it et al.}~\cite{casiulis2020velocity}, which describes a fluid of particles carrying ``spins'' (internal orientation vectors) coupled to their linear momentum. The Lagrangian for this system is given by:}
\begin{widetext}
\begin{equation}
\mathcal{L} = \sum_{i=1}^{N} \frac{m}{2} \dot{\vec{r}}_i^{\, 2} + \sum_{i=1}^{N} \frac{I}{2} \dot{\vec{S}}_i^{\,2} + \sum_{i=1}^{N} \vec{A}_i \cdot \dot{\vec{r}}_i - \frac{1}{2}\sum_{i \neq j} u(r_{ij}) + \frac{1}{2}J \sum_{i,j} g(r_{ij}) \vec{S}_i \cdot \vec{S}_j
\end{equation}
\end{widetext}
{Here, $\vec{r}_{i}$ represents the particle position, $\vec{S}_{i}$ denotes the internal spin (orientation), and $\vec{A}_{i}$ is a vector potential. The crucial term is $\vec{A}_i \cdot \dot{\vec{r}}_i$, which couples the particle's velocity to the vector potential. Casiulis {\it et al.}~chose a simple scenario where $\vec{A}_i = K \vec{S}_i$ and demonstrated that this model exhibits flocking-like behavior at lower temperatures. While there is ongoing debate regarding whether a moving cluster of particles can be strictly classified as a 'flock' in the traditional biological sense (e.g., bird flocks) \cite{cavagna2010scale, cavagna2019comment, casiulis2019reply}, the model nevertheless produces highly non-trivial and intriguing behavior. Although it remains to be established whether this framework faithfully mimics a collection of genuine active particles such as RTPs or ABPs, it provides a promising foundation for further theoretical development.}

{From a computational perspective, simulating such a system is non-trivial due to the velocity-dependent nature of the forces, meaning that standard molecular dynamics integrators like Velocity Verlet cannot be used directly \cite{leimkuhler2004simulating}. In recent work, we developed a symplectic integration scheme that preserves the geometric structure of phase space and guarantees long-term stability in energy conservation~\cite{bhattacharya2025thermostatting}. We also introduced a Nosé-Poincaré thermostat to sample the canonical ensemble correctly, which revealed that the canonical temperature in this class of models is markedly different and contains terms that depart from the usual kinetic energy (details in~\cite{bhattacharya2025thermostatting}). Developing future AH models to mimic ABPs or RTPs will unlock novel pathways to explore the coupling between random active forcing and long-wavelength phonons. Advantageously, the underlying Hamiltonian structure enables the calculation of both the standard dynamical matrix (the Hessian of the potential) and, crucially, the generalized dynamical matrix that includes velocity-dependent terms. Diagonalizing this matrix yields the eigenfrequencies (phonon spectrum) of the solid. We hypothesize that the spin-velocity coupling renormalizes the phonon frequencies, shifting them toward zero at small wavenumbers ($q \to 0$). This softening of the long-wavelength modes may provide a direct, mathematical explanation for the anomalous MWH fluctuations observed in standard active models. Furthermore, preliminary results suggest that this coupling excites breathing modes—collective compressions and expansions of particle clusters—similar to those observed in {\it Volvox} colonies~\cite{tan2022odd}.}

\vskip +0.1in
\noindent{\large\bf Chirality - A systematic Control Parameter for Coupling to Long-Wavelength Modes:}
{Since the instability of active solids arises from the coupling of activity to phonon modes, systematically modifying this coupling provides a powerful lever to tune their stability. ``Chirality''—the tendency of active particles to rotate or swim in circular trajectories—appears to offer exactly this control. While standard active particles, such as RTPs, typically destabilize solids, chiral active particles have been observed to suppress long-wavelength fluctuations. Crucially, recent simulations demonstrate that this mechanism can mitigate anomalous MWH fluctuations, thereby enabling true long-range positional order and the formation of ideal 2D crystals~\cite{kuroda2025long}. This presents a striking contrast to non-chiral active matter, where long-wavelength density fluctuations are enhanced rather than suppressed.}
 
{This creates a perfect setup to test our central hypothesis. If the coupling between random forces and phonons causes instability via an anomalous dispersion ($\omega(q) \sim q^{1.5}$), chirality must modify this behavior, either restoring the linear form ($\omega(q) \sim q$) or suppressing low-energy excitations by altering the scaling at small $q$. To systematically investigate this, we study chiral Run-and-Tumble Particles (cRTPs). Unlike conventional active particles that diffuse rotationally, cRTPs combine linear runs and discrete tumbles with a bias that enforces a circular trajectory. Adjusting the tumbling rate and bias allows one to continuously tune the system's degree of chirality.}

{Mapping the phase diagram of the active solid as a function of activity strength (\(f_{0}\)) and chirality (\(\Omega \)) would allow one to probe the transition line where the exponent governing the divergence of positional fluctuations changes. By employing the effective Hessian approach proposed in our recent work~\cite{dey2025enhanced}, one can extract the effective phonon dispersion relations as a function of changing chirality. Assuming that this effective Hessian framework extends universally to chiral systems, we expect to observe a systematic transition from sub-linear or anomalous spectra (characteristic of unstable active solids) to stable linear and super-linear dispersions as chirality stiffens the lattice. The implications of this study extend beyond fluctuation theorems to targeted material design. Specifically, because chirality suppresses long-wavelength modes, it likely alters how the solid responds to external stress—modes that are frequently responsible for the nucleation of shear bands, which drive localized, catastrophic failure.}

{We also hypothesize that chiral active solids will exhibit distinct yielding behaviors compared to their non-chiral counterparts. By tuning chirality, it may be possible to switch the failure mode from brittle (shear banding) to ductile (homogeneous flow). This concept paves the way for the design of ``mechanical circuit breakers'': materials that are stiff and brittle under certain conditions but can be made ductile and compliant by activating the chiral rotation of their constituent particles. For example, a material could be engineered to exhibit a stiff torsional response while remaining compliant in bending, purely by tuning the chirality of the active dopants~\cite{mallikarjun2023chiral}. Furthermore, incorporating chirality into the Active Hamiltonian framework represents a compelling direction. This unification would allow us to derive analytic expressions for how chirality modifies the generalized dynamical matrix, providing a rigorous theoretical basis for the suppression of fluctuations.}

\vskip +0.1in
\noindent{\large\bf The Oscillatory Shear Correspondence - Yielding and Memory:}
{Next, we address how to solidify the connection between active driving and mechanical deformation. We posit that the strong coupling of active forces to long-wavelength modes is phenomenologically identical to the effects of oscillatory shear. This equivalence offers a powerful avenue for understanding memory formation and yielding. Oscillatory shear imposes a global, time-dependent strain field $\gamma(t) = \gamma_0 \sin(\omega t)$, driving affine particle displacements interspersed with non-affine relaxations. In contrast, active driving introduces purely local, random force dipoles. Why these fundamentally different perturbations yield equivalent annealing effects remains a central open question.}

{The origin of this equivalence likely resides in how both mechanisms navigate the energy landscape to destabilize marginally stable states. Under shear deformation, plastic events occur along non-affine trajectories via saddle-node bifurcations~\cite{karmakar2010athermal}. Because these non-affine directions decompose into a few low-lying phonon modes dominated by quasi-localized excitations, oscillatory shear inevitably couples strongly to low-frequency modes. Strikingly, despite its localized nature, active driving similarly couples to these low-frequency phonons~\cite{dey2025enhanced}. This suggests that both phenomena are driven by the same fundamental coupling between the external excitation and low-frequency sectors. A key future objective is to determine whether active driving also induces a proliferation of quasi-localized modes. At large length scales, integrated active noise likely generates displacement fields that mathematically mimic macroscopic shear strain. A systematic investigation of this active-shear correspondence—using chirality to continuously tune long-wavelength couplings—presents a rigorous path to uncovering the physics behind this potentially universal phenomenon.}

\SK{A fascinating property of amorphous solids is their ability to encode memory. If a glass is subjected to oscillatory shear at a specific amplitude $\gamma_{train}$, it evolves into a state that ``remembers'' this amplitude. This memory can be read out by performing a strain sweep and observing a cusp in the diffusivity or stress at $\gamma_{train}$. If active driving is equivalent to shear, we should be able to:
\begin{enumerate}
\item {\bf Write with activity, read with shear:} Anneal a glass using active particles with a specific force $f_{train}$, then freeze the activity. Probe the passive solid with oscillatory shear. We predict a memory signature corresponding to the ``equivalent strain'' of that active force.
\item {\bf Write with shear, read with activity:} Shear anneal a glass, then turn on activity. We expect the mobility of the active particles to show a minimum when their effective force matches the training shear amplitude.
\end{enumerate}
}%
{Conventionally, memory formation is thought to require cyclic, deterministic driving. However, our preliminary results suggest that purely random active driving can also encode memory~\cite{sharma2025activity}, a finding supported by recent works showing that random driving encodes memory equally well~\cite{mungan2025self, chatterjee2026memory}. This strengthens the correspondence between active driving and oscillatory shear. Future exploration in these directions will provide crucial insights into how an amorphous solid navigates its potential energy landscape under the influence of either oscillatory shear (whether random or deterministic) or stochastic active forces~\cite{annurevMungan}. One important question to address is whether and how plasticity is modified by active forces. Does one observe the same saddle-node bifurcation during a plastic event in active glasses, or is it governed by a completely different instability? Furthermore, is it possible to exploit this equivalence to either enhance or suppress the response of solids under combined activity and oscillatory deformation? In another recent study~\cite{priya2025active}, we precisely explored this equivalence principle, leading to the discovery of physical phenomena that are typically absent in passive amorphous solids.}

{In Ref.~\cite{priya2025active}, we uncovered two pivotal insights into active glasses and brittle disordered solids. First, we identified a novel dynamical mechanism for plastic localization, where the interplay of shear-band propagation, internal activity, and external driving determines whether deformation localizes catastrophically or distributes across a network of plastic rearrangements. This allows for direct tuning of the yielding transition and the brittle-to-ductile response. Second, the study demonstrates that strain rate and active forcing serve as isomorphic driving variables. Introducing the effective control parameter $\dot{\gamma}f_0^m$ collapses the rheological response — including stress–strain profiles and yielding under simple shear or creep -- onto a universal master curve. This reveals a deep equivalence between mechanical and active driving, unifying active and passive amorphous rheology. Furthermore, the emergent strain-localization length scale obeys a specific scaling relation with this effective parameter, validated by simulation results. This approach explains the activity-induced reduction in localization length and further validates the proposed coupling of deformation and activity, offering new pathways to control active amorphous solids.}

\vskip +0.1in
\noindent{\large\bf Discussion:}
{Active systems exhibit phenomena that are absent in passive systems under equilibrium conditions. In this perspective, we have particularly highlighted enhanced MWH fluctuations, which can destabilize active solids even in 3D, or conversely, enable their stabilization via active forces. A key driver of these phenomena is the strong, non-perturbative coupling between spatially and temporally random active forces and long-wavelength density (phonon) modes. There is also a strong activity-oscillatory shear correspondence that highlights the importance of these long-wavelength modes in both driven and active systems. To obtain a deeper understanding of the non-equilibrium processes responsible for these collective phenomena, as well as the correspondence between active and driven systems, we propose the development of Active Hamiltonian (AH) models to serve as a reference framework. These models can map out pathways toward non-equilibrium active systems—ranging from chiral active matter to active gels—while elucidating the connection between active matter dynamics and the rheological response of passive systems under oscillatory shear deformation.}

\SK{The observed activity-oscillatory shear correspondence represents a potentially transformative conceptual breakthrough. If validated across a diverse set of systems as outlined in this perspective article, it would demonstrate that energy injection into amorphous solids, whether from internal active forces or external mechanical driving, follows universal principles governed by symmetry and conservation laws rather than microscopic details. This universality would place the field of active matter on firmer theoretical ground, analogous to how renormalization group theory unified our understanding of equilibrium critical phenomena despite the diversity of microscopic systems exhibiting criticality.}

{AH models offer a direct route to testing whether coupling active forces to long-wavelength modes drives non-equilibrium active behavior. Designing a Hamiltonian framework that replicates activity allows us to leverage equilibrium statistical mechanics while preserving non-conservative effects at macroscopic scales. Given that effective dynamical matrix calculations already rationalize key observations in active glasses and crystals, AH systems—which permit exact phonon spectra computation via generalized dynamical matrix diagonalization—represent a major methodological breakthrough. This approach may provide a mathematically rigorous explanation for the anomalous MWH fluctuations seen in standard active models. Finally, compelling parallels to living systems—ranging from multicellular Volvox colonies, wound healing, and cancer progression to the collective dynamics of ants and bird flocks—suggest that the principles governing active solids underpin fundamental organizational mechanisms across biology.}

\SK{The role of chirality as a control parameter for long-wavelength mode coupling represents an elegant experimental and theoretical test for the central hypothesis. The observation that chiral active particles suppress rather than enhance density fluctuations—potentially even restoring true long-range positional order in two dimensions—strongly supports the notion that the coupling of active driving with the phonon modes is probably the underlying universal mechanism in both chiral and non-chiral active solids. By systematically tuning the degree of chirality in Chiral Run-and-Tumble Particles and mapping the phase diagram as a function of activity strength and chirality, one can directly investigate how the phonon dispersion relation transitions from its anomalous super-linear behaviour, characteristic of unstable active solids, to a linear regime associated with marginally stable solids, and ultimately to a sub-linear regime that  characterizes stable active solids. Notably, the ability to shift failure modes from brittle to ductile by tuning chirality introduces the intriguing potential to engineer "mechanical circuit breakers"—materials whose mechanical properties can be dynamically adjusted by activating the chiral rotation of their constituent particles.}

\SK{Extension of this hypothesis to active gels will be very important to addresses a distinct but complementary aspect of the activity-structure coupling problem. Unlike dense active glasses where particles remain in close contact, gels are low-density, porous networks held together by attractive bonds. The demonstration that activity can accelerate structural evolution by orders of magnitude—achieving in hours what passive aging requires months to accomplish \cite{wei2023reconfiguration}—provides compelling evidence that active forces enable efficient exploration of the energy landscape even for less dense active solids like gels. Observation of an equivalence between active driving and oscillatory shear in active gels would be a significant scientific advance as validating this correspondence would allow prediction of active biological network rheology using established viscoelastic theories, with profound implications for understanding cytoskeletal dynamics, tissue mechanics, and the design of smart biomaterials with programmable stiffness and self-healing capabilities.}

{Crucially, the activity-shear correspondence acts as the conceptual linchpin connecting all elements of our roadmap. The hypothesis that active driving and mechanical deformation produce equivalent annealing due to identical long-wavelength mode couplings provides a unified view of non-equilibrium states. Our proposed cross-memory protocols—writing via activity and reading via shear—offer a definitive test: success would demonstrate that memory formation in amorphous systems can arise from purely stochastic active forces, bypassing the traditional requirement for cyclic, deterministic driving. From a broader perspective, verifying this universality has profound interdisciplinary implications. It introduces a rational design matrix for tailoring glassy materials; it illuminates how biological systems deploy active forcing to regulate structural remodeling and mechanical memory in development and disease; and it paves the way for smart, responsive soft matter that adapts autonomously to external loads.}

{This perspective highlights fundamental questions regarding how energy injection drives non-equilibrium states, how microscopic interactions dictate macroscopic mechanical response, and whether universal principles govern active matter. Systematically investigating long-wavelength modes across complementary models—from AH frameworks to chiral systems and network-forming gels—provides a unified lens for exploring dense active phases. This strategy not only uncovers universal large-scale emergent physics but also establishes a unified theoretical description of active solids, paving the way for the rational design of next-generation adaptive materials.}

\vskip +0.05in
\noindent{\large \bf Acknowledgments:}
\SK{We acknowledge funding by intramural funds at TIFR Hyderabad from the Department of Atomic Energy (DAE) under Project Identification No. RTI 4007. SK acknowledges the Swarna Jayanti Fellowship grants DST/SJF/PSA01/2018-19 and SB/SFJ/2019-20/05 from the Science and Engineering Research Board (SERB). Most of the computations are done using the HPC clusters procured using Swarna Jayanti Fellowship grants DST/SJF/PSA01/2018-19, SB/SFJ/2019-20/05 and Core Research Grant CRG/2019/005373. SK also acknowledges research support from MATRICES Grant MTR/2023/000079 from SERB. SK acknowledges generous financial support from Heinrich-Heine University (HHU) during SK's visit to HHU, where part of the article was written.}

\bibliography{references}

%merlin.mbs apsrev4-1.bst 2010-07-25 4.21a (PWD, AO, DPC) hacked
%Control: key (0)
%Control: author (8) initials jnrlst
%Control: editor formatted (1) identically to author
%Control: production of article title (-1) disabled
%Control: page (0) single
%Control: year (1) truncated
%Control: production of eprint (0) enabled
\begin{thebibliography}{47}%
\makeatletter
\providecommand \@ifxundefined [1]{%
 \@ifx{#1\undefined}
}%
\providecommand \@ifnum [1]{%
 \ifnum #1\expandafter \@firstoftwo
 \else \expandafter \@secondoftwo
 \fi
}%
\providecommand \@ifx [1]{%
 \ifx #1\expandafter \@firstoftwo
 \else \expandafter \@secondoftwo
 \fi
}%
\providecommand \natexlab [1]{#1}%
\providecommand \enquote  [1]{``#1''}%
\providecommand \bibnamefont  [1]{#1}%
\providecommand \bibfnamefont [1]{#1}%
\providecommand \citenamefont [1]{#1}%
\providecommand \href@noop [0]{\@secondoftwo}%
\providecommand \href [0]{\begingroup \@sanitize@url \@href}%
\providecommand \@href[1]{\@@startlink{#1}\@@href}%
\providecommand \@@href[1]{\endgroup#1\@@endlink}%
\providecommand \@sanitize@url [0]{\catcode `\\12\catcode `\$12\catcode
  `\&12\catcode `\#12\catcode `\^12\catcode `\_12\catcode `\%12\relax}%
\providecommand \@@startlink[1]{}%
\providecommand \@@endlink[0]{}%
\providecommand \url  [0]{\begingroup\@sanitize@url \@url }%
\providecommand \@url [1]{\endgroup\@href {#1}{\urlprefix }}%
\providecommand \urlprefix  [0]{URL }%
\providecommand \Eprint [0]{\href }%
\providecommand \doibase [0]{http://dx.doi.org/}%
\providecommand \selectlanguage [0]{\@gobble}%
\providecommand \bibinfo  [0]{\@secondoftwo}%
\providecommand \bibfield  [0]{\@secondoftwo}%
\providecommand \translation [1]{[#1]}%
\providecommand \BibitemOpen [0]{}%
\providecommand \bibitemStop [0]{}%
\providecommand \bibitemNoStop [0]{.\EOS\space}%
\providecommand \EOS [0]{\spacefactor3000\relax}%
\providecommand \BibitemShut  [1]{\csname bibitem#1\endcsname}%
\let\auto@bib@innerbib\@empty
%</preamble>
\bibitem [{\citenamefont {Ramaswamy}(2010)}]{ramaswamy2010mechanics}%
  \BibitemOpen
  \bibfield  {author} {\bibinfo {author} {\bibfnamefont {S.}~\bibnamefont
  {Ramaswamy}},\ }\href@noop {} {\bibfield  {journal} {\bibinfo  {journal}
  {Annu. Rev. Condens. Matter Phys.}\ }\textbf {\bibinfo {volume} {1}},\
  \bibinfo {pages} {323} (\bibinfo {year} {2010})}\BibitemShut {NoStop}%
\bibitem [{\citenamefont {Marchetti}\ \emph {et~al.}(2013)\citenamefont
  {Marchetti}, \citenamefont {Joanny}, \citenamefont {Ramaswamy}, \citenamefont
  {Liverpool}, \citenamefont {Prost}, \citenamefont {Rao},\ and\ \citenamefont
  {Simha}}]{marchetti2013hydrodynamics}%
  \BibitemOpen
  \bibfield  {author} {\bibinfo {author} {\bibfnamefont {M.~C.}\ \bibnamefont
  {Marchetti}}, \bibinfo {author} {\bibfnamefont {J.-F.}\ \bibnamefont
  {Joanny}}, \bibinfo {author} {\bibfnamefont {S.}~\bibnamefont {Ramaswamy}},
  \bibinfo {author} {\bibfnamefont {T.~B.}\ \bibnamefont {Liverpool}}, \bibinfo
  {author} {\bibfnamefont {J.}~\bibnamefont {Prost}}, \bibinfo {author}
  {\bibfnamefont {M.}~\bibnamefont {Rao}}, \ and\ \bibinfo {author}
  {\bibfnamefont {R.~A.}\ \bibnamefont {Simha}},\ }\href@noop {} {\bibfield
  {journal} {\bibinfo  {journal} {Reviews of modern physics}\ }\textbf
  {\bibinfo {volume} {85}},\ \bibinfo {pages} {1143} (\bibinfo {year}
  {2013})}\BibitemShut {NoStop}%
\bibitem [{\citenamefont {Palacci}\ \emph {et~al.}(2013)\citenamefont
  {Palacci}, \citenamefont {Sacanna}, \citenamefont {Steinberg}, \citenamefont
  {Pine},\ and\ \citenamefont {Chaikin}}]{palacci2013living}%
  \BibitemOpen
  \bibfield  {author} {\bibinfo {author} {\bibfnamefont {J.}~\bibnamefont
  {Palacci}}, \bibinfo {author} {\bibfnamefont {S.}~\bibnamefont {Sacanna}},
  \bibinfo {author} {\bibfnamefont {A.~P.}\ \bibnamefont {Steinberg}}, \bibinfo
  {author} {\bibfnamefont {D.~J.}\ \bibnamefont {Pine}}, \ and\ \bibinfo
  {author} {\bibfnamefont {P.~M.}\ \bibnamefont {Chaikin}},\ }\href@noop {}
  {\bibfield  {journal} {\bibinfo  {journal} {Science}\ }\textbf {\bibinfo
  {volume} {339}},\ \bibinfo {pages} {936} (\bibinfo {year}
  {2013})}\BibitemShut {NoStop}%
\bibitem [{\citenamefont {Vicsek}\ \emph {et~al.}(1995)\citenamefont {Vicsek},
  \citenamefont {Czir{\'o}k}, \citenamefont {Ben-Jacob}, \citenamefont
  {Cohen},\ and\ \citenamefont {Shochet}}]{vicsek1995novel}%
  \BibitemOpen
  \bibfield  {author} {\bibinfo {author} {\bibfnamefont {T.}~\bibnamefont
  {Vicsek}}, \bibinfo {author} {\bibfnamefont {A.}~\bibnamefont {Czir{\'o}k}},
  \bibinfo {author} {\bibfnamefont {E.}~\bibnamefont {Ben-Jacob}}, \bibinfo
  {author} {\bibfnamefont {I.}~\bibnamefont {Cohen}}, \ and\ \bibinfo {author}
  {\bibfnamefont {O.}~\bibnamefont {Shochet}},\ }\href@noop {} {\bibfield
  {journal} {\bibinfo  {journal} {Physical review letters}\ }\textbf {\bibinfo
  {volume} {75}},\ \bibinfo {pages} {1226} (\bibinfo {year}
  {1995})}\BibitemShut {NoStop}%
\bibitem [{\citenamefont {Ballerini}\ \emph {et~al.}(2008)\citenamefont
  {Ballerini}, \citenamefont {Cabibbo}, \citenamefont {Candelier},
  \citenamefont {Cavagna}, \citenamefont {Cisbani}, \citenamefont {Giardina},
  \citenamefont {Lecomte}, \citenamefont {Orlandi}, \citenamefont {Parisi},
  \citenamefont {Procaccini} \emph {et~al.}}]{ballerini2008interaction}%
  \BibitemOpen
  \bibfield  {author} {\bibinfo {author} {\bibfnamefont {M.}~\bibnamefont
  {Ballerini}}, \bibinfo {author} {\bibfnamefont {N.}~\bibnamefont {Cabibbo}},
  \bibinfo {author} {\bibfnamefont {R.}~\bibnamefont {Candelier}}, \bibinfo
  {author} {\bibfnamefont {A.}~\bibnamefont {Cavagna}}, \bibinfo {author}
  {\bibfnamefont {E.}~\bibnamefont {Cisbani}}, \bibinfo {author} {\bibfnamefont
  {I.}~\bibnamefont {Giardina}}, \bibinfo {author} {\bibfnamefont
  {V.}~\bibnamefont {Lecomte}}, \bibinfo {author} {\bibfnamefont
  {A.}~\bibnamefont {Orlandi}}, \bibinfo {author} {\bibfnamefont
  {G.}~\bibnamefont {Parisi}}, \bibinfo {author} {\bibfnamefont
  {A.}~\bibnamefont {Procaccini}},  \emph {et~al.},\ }\href@noop {} {\bibfield
  {journal} {\bibinfo  {journal} {Proceedings of the national academy of
  sciences}\ }\textbf {\bibinfo {volume} {105}},\ \bibinfo {pages} {1232}
  (\bibinfo {year} {2008})}\BibitemShut {NoStop}%
\bibitem [{\citenamefont {Toner}\ and\ \citenamefont
  {Tu}(1998)}]{toner1998flocks}%
  \BibitemOpen
  \bibfield  {author} {\bibinfo {author} {\bibfnamefont {J.}~\bibnamefont
  {Toner}}\ and\ \bibinfo {author} {\bibfnamefont {Y.}~\bibnamefont {Tu}},\
  }\href@noop {} {\bibfield  {journal} {\bibinfo  {journal} {Physical review
  E}\ }\textbf {\bibinfo {volume} {58}},\ \bibinfo {pages} {4828} (\bibinfo
  {year} {1998})}\BibitemShut {NoStop}%
\bibitem [{\citenamefont {Parry}\ \emph {et~al.}(2014)\citenamefont {Parry},
  \citenamefont {Surovtsev}, \citenamefont {Cabeen}, \citenamefont {O’hern},
  \citenamefont {Dufresne},\ and\ \citenamefont
  {Jacobs-Wagner}}]{parry2014bacterial}%
  \BibitemOpen
  \bibfield  {author} {\bibinfo {author} {\bibfnamefont {B.~R.}\ \bibnamefont
  {Parry}}, \bibinfo {author} {\bibfnamefont {I.~V.}\ \bibnamefont
  {Surovtsev}}, \bibinfo {author} {\bibfnamefont {M.~T.}\ \bibnamefont
  {Cabeen}}, \bibinfo {author} {\bibfnamefont {C.~S.}\ \bibnamefont
  {O’hern}}, \bibinfo {author} {\bibfnamefont {E.~R.}\ \bibnamefont
  {Dufresne}}, \ and\ \bibinfo {author} {\bibfnamefont {C.}~\bibnamefont
  {Jacobs-Wagner}},\ }\href@noop {} {\bibfield  {journal} {\bibinfo  {journal}
  {Cell}\ }\textbf {\bibinfo {volume} {156}},\ \bibinfo {pages} {183} (\bibinfo
  {year} {2014})}\BibitemShut {NoStop}%
\bibitem [{\citenamefont {Gravish}\ \emph {et~al.}(2015)\citenamefont
  {Gravish}, \citenamefont {Gold}, \citenamefont {Zangwill}, \citenamefont
  {Goodisman},\ and\ \citenamefont {Goldman}}]{gravish2015glass}%
  \BibitemOpen
  \bibfield  {author} {\bibinfo {author} {\bibfnamefont {N.}~\bibnamefont
  {Gravish}}, \bibinfo {author} {\bibfnamefont {G.}~\bibnamefont {Gold}},
  \bibinfo {author} {\bibfnamefont {A.}~\bibnamefont {Zangwill}}, \bibinfo
  {author} {\bibfnamefont {M.~A.}\ \bibnamefont {Goodisman}}, \ and\ \bibinfo
  {author} {\bibfnamefont {D.~I.}\ \bibnamefont {Goldman}},\ }\href@noop {}
  {\bibfield  {journal} {\bibinfo  {journal} {Soft matter}\ }\textbf {\bibinfo
  {volume} {11}},\ \bibinfo {pages} {6552} (\bibinfo {year}
  {2015})}\BibitemShut {NoStop}%
\bibitem [{\citenamefont {Angelini}\ \emph {et~al.}(2011)\citenamefont
  {Angelini}, \citenamefont {Hannezo}, \citenamefont {Trepat}, \citenamefont
  {Marquez}, \citenamefont {Fredberg},\ and\ \citenamefont
  {Weitz}}]{angelini2011glass}%
  \BibitemOpen
  \bibfield  {author} {\bibinfo {author} {\bibfnamefont {T.~E.}\ \bibnamefont
  {Angelini}}, \bibinfo {author} {\bibfnamefont {E.}~\bibnamefont {Hannezo}},
  \bibinfo {author} {\bibfnamefont {X.}~\bibnamefont {Trepat}}, \bibinfo
  {author} {\bibfnamefont {M.}~\bibnamefont {Marquez}}, \bibinfo {author}
  {\bibfnamefont {J.~J.}\ \bibnamefont {Fredberg}}, \ and\ \bibinfo {author}
  {\bibfnamefont {D.~A.}\ \bibnamefont {Weitz}},\ }\href@noop {} {\bibfield
  {journal} {\bibinfo  {journal} {Proceedings of the National Academy of
  Sciences}\ }\textbf {\bibinfo {volume} {108}},\ \bibinfo {pages} {4714}
  (\bibinfo {year} {2011})}\BibitemShut {NoStop}%
\bibitem [{\citenamefont {Dauchot}\ \emph {et~al.}(2005)\citenamefont
  {Dauchot}, \citenamefont {Marty},\ and\ \citenamefont
  {Biroli}}]{dauchot2005dynamical}%
  \BibitemOpen
  \bibfield  {author} {\bibinfo {author} {\bibfnamefont {O.}~\bibnamefont
  {Dauchot}}, \bibinfo {author} {\bibfnamefont {G.}~\bibnamefont {Marty}}, \
  and\ \bibinfo {author} {\bibfnamefont {G.}~\bibnamefont {Biroli}},\
  }\href@noop {} {\bibfield  {journal} {\bibinfo  {journal} {Physical review
  letters}\ }\textbf {\bibinfo {volume} {95}},\ \bibinfo {pages} {265701}
  (\bibinfo {year} {2005})}\BibitemShut {NoStop}%
\bibitem [{\citenamefont {Dreyfus}\ \emph {et~al.}(2005)\citenamefont
  {Dreyfus}, \citenamefont {Baudry}, \citenamefont {Roper}, \citenamefont
  {Fermigier}, \citenamefont {Stone},\ and\ \citenamefont
  {Bibette}}]{dreyfus2005microscopic}%
  \BibitemOpen
  \bibfield  {author} {\bibinfo {author} {\bibfnamefont {R.}~\bibnamefont
  {Dreyfus}}, \bibinfo {author} {\bibfnamefont {J.}~\bibnamefont {Baudry}},
  \bibinfo {author} {\bibfnamefont {M.~L.}\ \bibnamefont {Roper}}, \bibinfo
  {author} {\bibfnamefont {M.}~\bibnamefont {Fermigier}}, \bibinfo {author}
  {\bibfnamefont {H.~A.}\ \bibnamefont {Stone}}, \ and\ \bibinfo {author}
  {\bibfnamefont {J.}~\bibnamefont {Bibette}},\ }\href@noop {} {\bibfield
  {journal} {\bibinfo  {journal} {Nature}\ }\textbf {\bibinfo {volume} {437}},\
  \bibinfo {pages} {862} (\bibinfo {year} {2005})}\BibitemShut {NoStop}%
\bibitem [{\citenamefont {Takatori}\ and\ \citenamefont
  {Mandadapu}(2020)}]{takatori2020motility}%
  \BibitemOpen
  \bibfield  {author} {\bibinfo {author} {\bibfnamefont {S.~C.}\ \bibnamefont
  {Takatori}}\ and\ \bibinfo {author} {\bibfnamefont {K.~K.}\ \bibnamefont
  {Mandadapu}},\ }\href@noop {} {\bibfield  {journal} {\bibinfo  {journal}
  {arXiv preprint arXiv:2003.05618}\ } (\bibinfo {year} {2020})}\BibitemShut
  {NoStop}%
\bibitem [{\citenamefont {Mandal}\ \emph {et~al.}(2016)\citenamefont {Mandal},
  \citenamefont {Bhuyan}, \citenamefont {Rao},\ and\ \citenamefont
  {Dasgupta}}]{mandal2016active}%
  \BibitemOpen
  \bibfield  {author} {\bibinfo {author} {\bibfnamefont {R.}~\bibnamefont
  {Mandal}}, \bibinfo {author} {\bibfnamefont {P.~J.}\ \bibnamefont {Bhuyan}},
  \bibinfo {author} {\bibfnamefont {M.}~\bibnamefont {Rao}}, \ and\ \bibinfo
  {author} {\bibfnamefont {C.}~\bibnamefont {Dasgupta}},\ }\href@noop {}
  {\bibfield  {journal} {\bibinfo  {journal} {Soft Matter}\ }\textbf {\bibinfo
  {volume} {12}},\ \bibinfo {pages} {6268} (\bibinfo {year}
  {2016})}\BibitemShut {NoStop}%
\bibitem [{\citenamefont {Zhou}\ \emph {et~al.}(2009)\citenamefont {Zhou},
  \citenamefont {Trepat}, \citenamefont {Park}, \citenamefont {Lenormand},
  \citenamefont {Oliver}, \citenamefont {Mijailovich}, \citenamefont {Hardin},
  \citenamefont {Weitz}, \citenamefont {Butler},\ and\ \citenamefont
  {Fredberg}}]{zhou2009universal}%
  \BibitemOpen
  \bibfield  {author} {\bibinfo {author} {\bibfnamefont {E.}~\bibnamefont
  {Zhou}}, \bibinfo {author} {\bibfnamefont {X.}~\bibnamefont {Trepat}},
  \bibinfo {author} {\bibfnamefont {C.}~\bibnamefont {Park}}, \bibinfo {author}
  {\bibfnamefont {G.}~\bibnamefont {Lenormand}}, \bibinfo {author}
  {\bibfnamefont {M.}~\bibnamefont {Oliver}}, \bibinfo {author} {\bibfnamefont
  {S.}~\bibnamefont {Mijailovich}}, \bibinfo {author} {\bibfnamefont
  {C.}~\bibnamefont {Hardin}}, \bibinfo {author} {\bibfnamefont
  {D.}~\bibnamefont {Weitz}}, \bibinfo {author} {\bibfnamefont
  {J.}~\bibnamefont {Butler}}, \ and\ \bibinfo {author} {\bibfnamefont
  {J.}~\bibnamefont {Fredberg}},\ }\href@noop {} {\bibfield  {journal}
  {\bibinfo  {journal} {Proceedings of the National Academy of Sciences}\
  }\textbf {\bibinfo {volume} {106}},\ \bibinfo {pages} {10632} (\bibinfo
  {year} {2009})}\BibitemShut {NoStop}%
\bibitem [{\citenamefont {Berthier}\ and\ \citenamefont
  {Kurchan}(2013)}]{berthier2013non}%
  \BibitemOpen
  \bibfield  {author} {\bibinfo {author} {\bibfnamefont {L.}~\bibnamefont
  {Berthier}}\ and\ \bibinfo {author} {\bibfnamefont {J.}~\bibnamefont
  {Kurchan}},\ }\href@noop {} {\bibfield  {journal} {\bibinfo  {journal}
  {Nature Physics}\ }\textbf {\bibinfo {volume} {9}},\ \bibinfo {pages} {310}
  (\bibinfo {year} {2013})}\BibitemShut {NoStop}%
\bibitem [{\citenamefont {Berthier}(2014)}]{berthier2014nonequilibrium}%
  \BibitemOpen
  \bibfield  {author} {\bibinfo {author} {\bibfnamefont {L.}~\bibnamefont
  {Berthier}},\ }\href@noop {} {\bibfield  {journal} {\bibinfo  {journal}
  {Physical review letters}\ }\textbf {\bibinfo {volume} {112}},\ \bibinfo
  {pages} {220602} (\bibinfo {year} {2014})}\BibitemShut {NoStop}%
\bibitem [{\citenamefont {Fodor}\ \emph {et~al.}(2016)\citenamefont {Fodor},
  \citenamefont {Nardini}, \citenamefont {Cates}, \citenamefont {Tailleur},
  \citenamefont {Visco},\ and\ \citenamefont {Van~Wijland}}]{fodor2016far}%
  \BibitemOpen
  \bibfield  {author} {\bibinfo {author} {\bibfnamefont {{\'E}.}~\bibnamefont
  {Fodor}}, \bibinfo {author} {\bibfnamefont {C.}~\bibnamefont {Nardini}},
  \bibinfo {author} {\bibfnamefont {M.~E.}\ \bibnamefont {Cates}}, \bibinfo
  {author} {\bibfnamefont {J.}~\bibnamefont {Tailleur}}, \bibinfo {author}
  {\bibfnamefont {P.}~\bibnamefont {Visco}}, \ and\ \bibinfo {author}
  {\bibfnamefont {F.}~\bibnamefont {Van~Wijland}},\ }\href@noop {} {\bibfield
  {journal} {\bibinfo  {journal} {Physical review letters}\ }\textbf {\bibinfo
  {volume} {117}},\ \bibinfo {pages} {038103} (\bibinfo {year}
  {2016})}\BibitemShut {NoStop}%
\bibitem [{\citenamefont {Loi}\ \emph {et~al.}(2008)\citenamefont {Loi},
  \citenamefont {Mossa},\ and\ \citenamefont {Cugliandolo}}]{loi2008effective}%
  \BibitemOpen
  \bibfield  {author} {\bibinfo {author} {\bibfnamefont {D.}~\bibnamefont
  {Loi}}, \bibinfo {author} {\bibfnamefont {S.}~\bibnamefont {Mossa}}, \ and\
  \bibinfo {author} {\bibfnamefont {L.~F.}\ \bibnamefont {Cugliandolo}},\
  }\href@noop {} {\bibfield  {journal} {\bibinfo  {journal} {Physical Review
  E—Statistical, Nonlinear, and Soft Matter Physics}\ }\textbf {\bibinfo
  {volume} {77}},\ \bibinfo {pages} {051111} (\bibinfo {year}
  {2008})}\BibitemShut {NoStop}%
\bibitem [{\citenamefont {Nandi}\ \emph {et~al.}(2018)\citenamefont {Nandi},
  \citenamefont {Mandal}, \citenamefont {Bhuyan}, \citenamefont {Dasgupta},
  \citenamefont {Rao},\ and\ \citenamefont {Gov}}]{nandi2018random}%
  \BibitemOpen
  \bibfield  {author} {\bibinfo {author} {\bibfnamefont {S.~K.}\ \bibnamefont
  {Nandi}}, \bibinfo {author} {\bibfnamefont {R.}~\bibnamefont {Mandal}},
  \bibinfo {author} {\bibfnamefont {P.~J.}\ \bibnamefont {Bhuyan}}, \bibinfo
  {author} {\bibfnamefont {C.}~\bibnamefont {Dasgupta}}, \bibinfo {author}
  {\bibfnamefont {M.}~\bibnamefont {Rao}}, \ and\ \bibinfo {author}
  {\bibfnamefont {N.~S.}\ \bibnamefont {Gov}},\ }\href@noop {} {\bibfield
  {journal} {\bibinfo  {journal} {Proceedings of the National Academy of
  Sciences}\ }\textbf {\bibinfo {volume} {115}},\ \bibinfo {pages} {7688}
  (\bibinfo {year} {2018})}\BibitemShut {NoStop}%
\bibitem [{\citenamefont {Ni}\ \emph {et~al.}(2013)\citenamefont {Ni},
  \citenamefont {Stuart},\ and\ \citenamefont {Dijkstra}}]{ni2013pushing}%
  \BibitemOpen
  \bibfield  {author} {\bibinfo {author} {\bibfnamefont {R.}~\bibnamefont
  {Ni}}, \bibinfo {author} {\bibfnamefont {M.~A.~C.}\ \bibnamefont {Stuart}}, \
  and\ \bibinfo {author} {\bibfnamefont {M.}~\bibnamefont {Dijkstra}},\
  }\href@noop {} {\bibfield  {journal} {\bibinfo  {journal} {Nature
  communications}\ }\textbf {\bibinfo {volume} {4}},\ \bibinfo {pages} {2704}
  (\bibinfo {year} {2013})}\BibitemShut {NoStop}%
\bibitem [{\citenamefont {Berthier}\ \emph {et~al.}(2019)\citenamefont
  {Berthier}, \citenamefont {Flenner},\ and\ \citenamefont
  {Szamel}}]{berthier2019glassy}%
  \BibitemOpen
  \bibfield  {author} {\bibinfo {author} {\bibfnamefont {L.}~\bibnamefont
  {Berthier}}, \bibinfo {author} {\bibfnamefont {E.}~\bibnamefont {Flenner}}, \
  and\ \bibinfo {author} {\bibfnamefont {G.}~\bibnamefont {Szamel}},\
  }\href@noop {} {\bibfield  {journal} {\bibinfo  {journal} {The Journal of
  chemical physics}\ }\textbf {\bibinfo {volume} {150}} (\bibinfo {year}
  {2019})}\BibitemShut {NoStop}%
\bibitem [{\citenamefont {Paul}\ \emph {et~al.}(2023)\citenamefont {Paul},
  \citenamefont {Mutneja}, \citenamefont {Nandi},\ and\ \citenamefont
  {Karmakar}}]{paul2023dynamical}%
  \BibitemOpen
  \bibfield  {author} {\bibinfo {author} {\bibfnamefont {K.}~\bibnamefont
  {Paul}}, \bibinfo {author} {\bibfnamefont {A.}~\bibnamefont {Mutneja}},
  \bibinfo {author} {\bibfnamefont {S.~K.}\ \bibnamefont {Nandi}}, \ and\
  \bibinfo {author} {\bibfnamefont {S.}~\bibnamefont {Karmakar}},\ }\href@noop
  {} {\bibfield  {journal} {\bibinfo  {journal} {Proceedings of the National
  Academy of Sciences}\ }\textbf {\bibinfo {volume} {120}},\ \bibinfo {pages}
  {e2217073120} (\bibinfo {year} {2023})}\BibitemShut {NoStop}%
\bibitem [{\citenamefont {Chatterjee}\ \emph
  {et~al.}(2026{\natexlab{a}})\citenamefont {Chatterjee}, \citenamefont
  {Adhikari},\ and\ \citenamefont {Karmakar}}]{RoniNatComm2026}%
  \BibitemOpen
  \bibfield  {author} {\bibinfo {author} {\bibfnamefont {R.}~\bibnamefont
  {Chatterjee}}, \bibinfo {author} {\bibfnamefont {M.}~\bibnamefont
  {Adhikari}}, \ and\ \bibinfo {author} {\bibfnamefont {S.}~\bibnamefont
  {Karmakar}},\ }\href {\doibase 10.1038/s41467-026-71157-w} {\bibfield
  {journal} {\bibinfo  {journal} {Nature Communications}\ }\textbf {\bibinfo
  {volume} {17}},\ \bibinfo {pages} {4506} (\bibinfo {year}
  {2026}{\natexlab{a}})}\BibitemShut {NoStop}%
\bibitem [{\citenamefont {Hohenberg}(1967)}]{hohenberg1967existence}%
  \BibitemOpen
  \bibfield  {author} {\bibinfo {author} {\bibfnamefont {P.~C.}\ \bibnamefont
  {Hohenberg}},\ }\href@noop {} {\bibfield  {journal} {\bibinfo  {journal}
  {Physical Review}\ }\textbf {\bibinfo {volume} {158}},\ \bibinfo {pages}
  {383} (\bibinfo {year} {1967})}\BibitemShut {NoStop}%
\bibitem [{\citenamefont {Mermin}\ and\ \citenamefont
  {Wagner}(1966)}]{mermin1966absence}%
  \BibitemOpen
  \bibfield  {author} {\bibinfo {author} {\bibfnamefont {N.~D.}\ \bibnamefont
  {Mermin}}\ and\ \bibinfo {author} {\bibfnamefont {H.}~\bibnamefont
  {Wagner}},\ }\href@noop {} {\bibfield  {journal} {\bibinfo  {journal}
  {Physical review letters}\ }\textbf {\bibinfo {volume} {17}},\ \bibinfo
  {pages} {1133} (\bibinfo {year} {1966})}\BibitemShut {NoStop}%
\bibitem [{\citenamefont {Mermin}(1968)}]{mermin1968crystalline}%
  \BibitemOpen
  \bibfield  {author} {\bibinfo {author} {\bibfnamefont {N.~D.}\ \bibnamefont
  {Mermin}},\ }\href@noop {} {\bibfield  {journal} {\bibinfo  {journal}
  {Physical review}\ }\textbf {\bibinfo {volume} {176}},\ \bibinfo {pages}
  {250} (\bibinfo {year} {1968})}\BibitemShut {NoStop}%
\bibitem [{\citenamefont {Jancovici}(1967)}]{jancovici1967infinite}%
  \BibitemOpen
  \bibfield  {author} {\bibinfo {author} {\bibfnamefont {B.}~\bibnamefont
  {Jancovici}},\ }\href@noop {} {\bibfield  {journal} {\bibinfo  {journal}
  {Physical Review Letters}\ }\textbf {\bibinfo {volume} {19}},\ \bibinfo
  {pages} {20} (\bibinfo {year} {1967})}\BibitemShut {NoStop}%
\bibitem [{\citenamefont {Dey}\ \emph {et~al.}(2025)\citenamefont {Dey},
  \citenamefont {Bhattacharya},\ and\ \citenamefont
  {Karmakar}}]{dey2025enhanced}%
  \BibitemOpen
  \bibfield  {author} {\bibinfo {author} {\bibfnamefont {S.}~\bibnamefont
  {Dey}}, \bibinfo {author} {\bibfnamefont {A.}~\bibnamefont {Bhattacharya}}, \
  and\ \bibinfo {author} {\bibfnamefont {S.}~\bibnamefont {Karmakar}},\
  }\href@noop {} {\bibfield  {journal} {\bibinfo  {journal} {Nature
  Communications}\ }\textbf {\bibinfo {volume} {16}},\ \bibinfo {pages} {5498}
  (\bibinfo {year} {2025})}\BibitemShut {NoStop}%
\bibitem [{\citenamefont {Galliano}\ \emph {et~al.}(2023)\citenamefont
  {Galliano}, \citenamefont {Cates},\ and\ \citenamefont
  {Berthier}}]{galliano2023two}%
  \BibitemOpen
  \bibfield  {author} {\bibinfo {author} {\bibfnamefont {L.}~\bibnamefont
  {Galliano}}, \bibinfo {author} {\bibfnamefont {M.~E.}\ \bibnamefont {Cates}},
  \ and\ \bibinfo {author} {\bibfnamefont {L.}~\bibnamefont {Berthier}},\
  }\href@noop {} {\bibfield  {journal} {\bibinfo  {journal} {Physical Review
  Letters}\ }\textbf {\bibinfo {volume} {131}},\ \bibinfo {pages} {047101}
  (\bibinfo {year} {2023})}\BibitemShut {NoStop}%
\bibitem [{\citenamefont {Kuroda}\ \emph {et~al.}(2025)\citenamefont {Kuroda},
  \citenamefont {Kawasaki},\ and\ \citenamefont {Miyazaki}}]{kuroda2025long}%
  \BibitemOpen
  \bibfield  {author} {\bibinfo {author} {\bibfnamefont {Y.}~\bibnamefont
  {Kuroda}}, \bibinfo {author} {\bibfnamefont {T.}~\bibnamefont {Kawasaki}}, \
  and\ \bibinfo {author} {\bibfnamefont {K.}~\bibnamefont {Miyazaki}},\
  }\href@noop {} {\bibfield  {journal} {\bibinfo  {journal} {Physical Review
  Research}\ }\textbf {\bibinfo {volume} {7}},\ \bibinfo {pages} {L012048}
  (\bibinfo {year} {2025})}\BibitemShut {NoStop}%
\bibitem [{\citenamefont {Sharma}\ and\ \citenamefont
  {Karmakar}(2025)}]{sharma2025activity}%
  \BibitemOpen
  \bibfield  {author} {\bibinfo {author} {\bibfnamefont {R.}~\bibnamefont
  {Sharma}}\ and\ \bibinfo {author} {\bibfnamefont {S.}~\bibnamefont
  {Karmakar}},\ }\href@noop {} {\bibfield  {journal} {\bibinfo  {journal}
  {Nature Physics}\ }\textbf {\bibinfo {volume} {21}},\ \bibinfo {pages} {253}
  (\bibinfo {year} {2025})}\BibitemShut {NoStop}%
\bibitem [{\citenamefont {Goswami}\ \emph {et~al.}(2025)\citenamefont
  {Goswami}, \citenamefont {Shivashankar},\ and\ \citenamefont
  {Sastry}}]{goswami2025yielding}%
  \BibitemOpen
  \bibfield  {author} {\bibinfo {author} {\bibfnamefont {Y.}~\bibnamefont
  {Goswami}}, \bibinfo {author} {\bibfnamefont {G.}~\bibnamefont
  {Shivashankar}}, \ and\ \bibinfo {author} {\bibfnamefont {S.}~\bibnamefont
  {Sastry}},\ }\href@noop {} {\bibfield  {journal} {\bibinfo  {journal} {Nature
  Physics}\ }\textbf {\bibinfo {volume} {21}},\ \bibinfo {pages} {817}
  (\bibinfo {year} {2025})}\BibitemShut {NoStop}%
\bibitem [{\citenamefont {L{\o}land~Bore}\ \emph {et~al.}(2016)\citenamefont
  {L{\o}land~Bore}, \citenamefont {Schindler}, \citenamefont {Nguyen Thu~Lam},
  \citenamefont {Bertin},\ and\ \citenamefont {Dauchot}}]{loland2016coupling}%
  \BibitemOpen
  \bibfield  {author} {\bibinfo {author} {\bibfnamefont {S.}~\bibnamefont
  {L{\o}land~Bore}}, \bibinfo {author} {\bibfnamefont {M.}~\bibnamefont
  {Schindler}}, \bibinfo {author} {\bibfnamefont {K.-D.}\ \bibnamefont {Nguyen
  Thu~Lam}}, \bibinfo {author} {\bibfnamefont {E.}~\bibnamefont {Bertin}}, \
  and\ \bibinfo {author} {\bibfnamefont {O.}~\bibnamefont {Dauchot}},\
  }\href@noop {} {\bibfield  {journal} {\bibinfo  {journal} {Journal of
  Statistical Mechanics: Theory and Experiment}\ }\textbf {\bibinfo {volume}
  {2016}},\ \bibinfo {pages} {033305} (\bibinfo {year} {2016})}\BibitemShut
  {NoStop}%
\bibitem [{\citenamefont {Casiulis}\ \emph {et~al.}(2020)\citenamefont
  {Casiulis}, \citenamefont {Tarzia}, \citenamefont {Cugliandolo},\ and\
  \citenamefont {Dauchot}}]{casiulis2020velocity}%
  \BibitemOpen
  \bibfield  {author} {\bibinfo {author} {\bibfnamefont {M.}~\bibnamefont
  {Casiulis}}, \bibinfo {author} {\bibfnamefont {M.}~\bibnamefont {Tarzia}},
  \bibinfo {author} {\bibfnamefont {L.~F.}\ \bibnamefont {Cugliandolo}}, \ and\
  \bibinfo {author} {\bibfnamefont {O.}~\bibnamefont {Dauchot}},\ }\href@noop
  {} {\bibfield  {journal} {\bibinfo  {journal} {Physical Review Letters}\
  }\textbf {\bibinfo {volume} {124}},\ \bibinfo {pages} {198001} (\bibinfo
  {year} {2020})}\BibitemShut {NoStop}%
\bibitem [{\citenamefont {Cavagna}\ \emph {et~al.}(2010)\citenamefont
  {Cavagna}, \citenamefont {Cimarelli}, \citenamefont {Giardina}, \citenamefont
  {Parisi}, \citenamefont {Santagati}, \citenamefont {Stefanini},\ and\
  \citenamefont {Viale}}]{cavagna2010scale}%
  \BibitemOpen
  \bibfield  {author} {\bibinfo {author} {\bibfnamefont {A.}~\bibnamefont
  {Cavagna}}, \bibinfo {author} {\bibfnamefont {A.}~\bibnamefont {Cimarelli}},
  \bibinfo {author} {\bibfnamefont {I.}~\bibnamefont {Giardina}}, \bibinfo
  {author} {\bibfnamefont {G.}~\bibnamefont {Parisi}}, \bibinfo {author}
  {\bibfnamefont {R.}~\bibnamefont {Santagati}}, \bibinfo {author}
  {\bibfnamefont {F.}~\bibnamefont {Stefanini}}, \ and\ \bibinfo {author}
  {\bibfnamefont {M.}~\bibnamefont {Viale}},\ }\href@noop {} {\bibfield
  {journal} {\bibinfo  {journal} {Proceedings of the National Academy of
  Sciences}\ }\textbf {\bibinfo {volume} {107}},\ \bibinfo {pages} {11865}
  (\bibinfo {year} {2010})}\BibitemShut {NoStop}%
\bibitem [{\citenamefont {Cavagna}\ \emph {et~al.}(2019)\citenamefont
  {Cavagna}, \citenamefont {Giardina},\ and\ \citenamefont
  {Viale}}]{cavagna2019comment}%
  \BibitemOpen
  \bibfield  {author} {\bibinfo {author} {\bibfnamefont {A.}~\bibnamefont
  {Cavagna}}, \bibinfo {author} {\bibfnamefont {I.}~\bibnamefont {Giardina}}, \
  and\ \bibinfo {author} {\bibfnamefont {M.}~\bibnamefont {Viale}},\
  }\href@noop {} {\bibfield  {journal} {\bibinfo  {journal} {arXiv preprint
  arXiv:1912.07056}\ } (\bibinfo {year} {2019})}\BibitemShut {NoStop}%
\bibitem [{\citenamefont {Casiulis}\ \emph {et~al.}(2019)\citenamefont
  {Casiulis}, \citenamefont {Tarzia}, \citenamefont {Cugliandolo},\ and\
  \citenamefont {Dauchot}}]{casiulis2019reply}%
  \BibitemOpen
  \bibfield  {author} {\bibinfo {author} {\bibfnamefont {M.}~\bibnamefont
  {Casiulis}}, \bibinfo {author} {\bibfnamefont {M.}~\bibnamefont {Tarzia}},
  \bibinfo {author} {\bibfnamefont {L.~F.}\ \bibnamefont {Cugliandolo}}, \ and\
  \bibinfo {author} {\bibfnamefont {O.}~\bibnamefont {Dauchot}},\ }\href@noop
  {} {\bibfield  {journal} {\bibinfo  {journal} {arXiv preprint
  arXiv:1912.09202}\ } (\bibinfo {year} {2019})}\BibitemShut {NoStop}%
\bibitem [{\citenamefont {Leimkuhler}\ and\ \citenamefont
  {Reich}(2004)}]{leimkuhler2004simulating}%
  \BibitemOpen
  \bibfield  {author} {\bibinfo {author} {\bibfnamefont {B.}~\bibnamefont
  {Leimkuhler}}\ and\ \bibinfo {author} {\bibfnamefont {S.}~\bibnamefont
  {Reich}},\ }\href@noop {} {\emph {\bibinfo {title} {Simulating hamiltonian
  dynamics}}},\ \bibinfo {number} {14}\ (\bibinfo  {publisher} {Cambridge
  university press},\ \bibinfo {year} {2004})\BibitemShut {NoStop}%
\bibitem [{\citenamefont {Bhattacharya}\ \emph {et~al.}(2025)\citenamefont
  {Bhattacharya}, \citenamefont {Horbach},\ and\ \citenamefont
  {Karmakar}}]{bhattacharya2025thermostatting}%
  \BibitemOpen
  \bibfield  {author} {\bibinfo {author} {\bibfnamefont {A.}~\bibnamefont
  {Bhattacharya}}, \bibinfo {author} {\bibfnamefont {J.}~\bibnamefont
  {Horbach}}, \ and\ \bibinfo {author} {\bibfnamefont {S.}~\bibnamefont
  {Karmakar}},\ }\href@noop {} {\bibfield  {journal} {\bibinfo  {journal}
  {Physical Review E}\ }\textbf {\bibinfo {volume} {111}},\ \bibinfo {pages}
  {015429} (\bibinfo {year} {2025})}\BibitemShut {NoStop}%
\bibitem [{\citenamefont {Tan}\ \emph {et~al.}(2022)\citenamefont {Tan},
  \citenamefont {Mietke}, \citenamefont {Li}, \citenamefont {Chen},
  \citenamefont {Higinbotham}, \citenamefont {Foster}, \citenamefont {Gokhale},
  \citenamefont {Dunkel},\ and\ \citenamefont {Fakhri}}]{tan2022odd}%
  \BibitemOpen
  \bibfield  {author} {\bibinfo {author} {\bibfnamefont {T.~H.}\ \bibnamefont
  {Tan}}, \bibinfo {author} {\bibfnamefont {A.}~\bibnamefont {Mietke}},
  \bibinfo {author} {\bibfnamefont {J.}~\bibnamefont {Li}}, \bibinfo {author}
  {\bibfnamefont {Y.}~\bibnamefont {Chen}}, \bibinfo {author} {\bibfnamefont
  {H.}~\bibnamefont {Higinbotham}}, \bibinfo {author} {\bibfnamefont {P.~J.}\
  \bibnamefont {Foster}}, \bibinfo {author} {\bibfnamefont {S.}~\bibnamefont
  {Gokhale}}, \bibinfo {author} {\bibfnamefont {J.}~\bibnamefont {Dunkel}}, \
  and\ \bibinfo {author} {\bibfnamefont {N.}~\bibnamefont {Fakhri}},\
  }\href@noop {} {\bibfield  {journal} {\bibinfo  {journal} {Nature}\ }\textbf
  {\bibinfo {volume} {607}},\ \bibinfo {pages} {287} (\bibinfo {year}
  {2022})}\BibitemShut {NoStop}%
\bibitem [{\citenamefont {Mallikarjun}\ and\ \citenamefont
  {Pal}(2023)}]{mallikarjun2023chiral}%
  \BibitemOpen
  \bibfield  {author} {\bibinfo {author} {\bibfnamefont {R.}~\bibnamefont
  {Mallikarjun}}\ and\ \bibinfo {author} {\bibfnamefont {A.}~\bibnamefont
  {Pal}},\ }\href@noop {} {\bibfield  {journal} {\bibinfo  {journal} {Physica
  A: Statistical Mechanics and its Applications}\ }\textbf {\bibinfo {volume}
  {622}},\ \bibinfo {pages} {128821} (\bibinfo {year} {2023})}\BibitemShut
  {NoStop}%
\bibitem [{\citenamefont {Karmakar}\ \emph {et~al.}(2010)\citenamefont
  {Karmakar}, \citenamefont {Lerner},\ and\ \citenamefont
  {Procaccia}}]{karmakar2010athermal}%
  \BibitemOpen
  \bibfield  {author} {\bibinfo {author} {\bibfnamefont {S.}~\bibnamefont
  {Karmakar}}, \bibinfo {author} {\bibfnamefont {E.}~\bibnamefont {Lerner}}, \
  and\ \bibinfo {author} {\bibfnamefont {I.}~\bibnamefont {Procaccia}},\
  }\href@noop {} {\bibfield  {journal} {\bibinfo  {journal} {Physical Review
  E—Statistical, Nonlinear, and Soft Matter Physics}\ }\textbf {\bibinfo
  {volume} {82}},\ \bibinfo {pages} {026105} (\bibinfo {year}
  {2010})}\BibitemShut {NoStop}%
\bibitem [{\citenamefont {Mungan}\ \emph {et~al.}(2025)\citenamefont {Mungan},
  \citenamefont {Kumar}, \citenamefont {Patinet},\ and\ \citenamefont
  {Vandembroucq}}]{mungan2025self}%
  \BibitemOpen
  \bibfield  {author} {\bibinfo {author} {\bibfnamefont {M.}~\bibnamefont
  {Mungan}}, \bibinfo {author} {\bibfnamefont {D.}~\bibnamefont {Kumar}},
  \bibinfo {author} {\bibfnamefont {S.}~\bibnamefont {Patinet}}, \ and\
  \bibinfo {author} {\bibfnamefont {D.}~\bibnamefont {Vandembroucq}},\
  }\href@noop {} {\bibfield  {journal} {\bibinfo  {journal} {Physical Review
  Letters}\ }\textbf {\bibinfo {volume} {134}},\ \bibinfo {pages} {178203}
  (\bibinfo {year} {2025})}\BibitemShut {NoStop}%
\bibitem [{\citenamefont {Chatterjee}\ \emph
  {et~al.}(2026{\natexlab{b}})\citenamefont {Chatterjee}, \citenamefont
  {Karamkar}, \citenamefont {Mungan},\ and\ \citenamefont
  {Vandembroucq}}]{chatterjee2026memory}%
  \BibitemOpen
  \bibfield  {author} {\bibinfo {author} {\bibfnamefont {R.}~\bibnamefont
  {Chatterjee}}, \bibinfo {author} {\bibfnamefont {S.}~\bibnamefont
  {Karamkar}}, \bibinfo {author} {\bibfnamefont {M.}~\bibnamefont {Mungan}}, \
  and\ \bibinfo {author} {\bibfnamefont {D.}~\bibnamefont {Vandembroucq}},\
  }\href@noop {} {\bibfield  {journal} {\bibinfo  {journal} {New Journal of
  Physics}\ }\textbf {\bibinfo {volume} {28}},\ \bibinfo {pages} {044401}
  (\bibinfo {year} {2026}{\natexlab{b}})}\BibitemShut {NoStop}%
\bibitem [{\citenamefont {Mungan}\ \emph {et~al.}(2026)\citenamefont {Mungan},
  \citenamefont {Clément}, \citenamefont {Vandembroucq},\ and\ \citenamefont
  {Sastry}}]{annurevMungan}%
  \BibitemOpen
  \bibfield  {author} {\bibinfo {author} {\bibfnamefont {M.}~\bibnamefont
  {Mungan}}, \bibinfo {author} {\bibfnamefont {E.}~\bibnamefont {Clément}},
  \bibinfo {author} {\bibfnamefont {D.}~\bibnamefont {Vandembroucq}}, \ and\
  \bibinfo {author} {\bibfnamefont {S.}~\bibnamefont {Sastry}},\ }\href
  {\doibase https://doi.org/10.1146/annurev-conmatphys-082225-051908}
  {\bibfield  {journal} {\bibinfo  {journal} {Annual Review of Condensed Matter
  Physics}\ }\textbf {\bibinfo {volume} {17}},\ \bibinfo {pages} {207}
  (\bibinfo {year} {2026})}\BibitemShut {NoStop}%
\bibitem [{\citenamefont {Priya}\ \emph {et~al.}(2025)\citenamefont {Priya},
  \citenamefont {Horbach},\ and\ \citenamefont {Karmakar}}]{priya2025active}%
  \BibitemOpen
  \bibfield  {author} {\bibinfo {author} {\bibfnamefont {R.}~\bibnamefont
  {Priya}}, \bibinfo {author} {\bibfnamefont {J.}~\bibnamefont {Horbach}}, \
  and\ \bibinfo {author} {\bibfnamefont {S.}~\bibnamefont {Karmakar}},\
  }\href@noop {} {\bibfield  {journal} {\bibinfo  {journal} {Nature
  Communications 2026 arXiv preprint arXiv:2508.06260}\ } (\bibinfo {year}
  {2025})}\BibitemShut {NoStop}%
\bibitem [{\citenamefont {Wei}\ \emph {et~al.}(2023)\citenamefont {Wei},
  \citenamefont {Ben~Zion},\ and\ \citenamefont
  {Dauchot}}]{wei2023reconfiguration}%
  \BibitemOpen
  \bibfield  {author} {\bibinfo {author} {\bibfnamefont {M.}~\bibnamefont
  {Wei}}, \bibinfo {author} {\bibfnamefont {M.~Y.}\ \bibnamefont {Ben~Zion}}, \
  and\ \bibinfo {author} {\bibfnamefont {O.}~\bibnamefont {Dauchot}},\
  }\href@noop {} {\bibfield  {journal} {\bibinfo  {journal} {Physical Review
  Letters}\ }\textbf {\bibinfo {volume} {131}},\ \bibinfo {pages} {018301}
  (\bibinfo {year} {2023})}\BibitemShut {NoStop}%
\end{thebibliography}%

\end{document}